\documentclass[proceedings]{stacs}
\stacsheading{2009}{505--516}{Freiburg}
\firstpageno{505}

\newcommand{\eq}{\!=\!}
\renewcommand{\vec}[1]{{\bf{}#1}}

\newcommand{\half}{\ensuremath{\frac{1}{2}}}
\newcommand{\col}{\ensuremath{\!:\!}}

\newcommand{\binset}{\ensuremath{\{0,1\}}}

\newcommand{\support}{\operatorname{support}}

\newcommand{\AND}{\operatorname{AND}}
\newcommand{\OR}{\operatorname{OR}}

\newcommand{\VV}{\operatorname{V}}
\newcommand{\DD}{\operatorname{D}}
\newcommand{\II}{\operatorname{I}}
\newcommand{\HH}{\operatorname{H}}

\newcommand{\icost}{\operatorname{icost}}
\newcommand{\IC}{\operatorname{IC}}

\newcommand{\eps}{\varepsilon}

\newcommand{\proofsketch}{\proof[Proof Sketch]}

\begin{document}
\title[NIH-Multi-Party Information Complexity of Disjointness]
    {Asymptotically Optimal Lower Bounds on the NIH-Multi-Party
    Information Complexity of the {\boldmath $\AND$-Function} and Disjointness}
\author[AG]{A. Gronemeier}{Andr\'{e} Gronemeier}
\address[AG]{Technische Universit\"{a}t Dortmund, Lehrstuhl Informatik 2,
44227 Dortmund, Germany}
\email{andre.gronemeier@cs.uni-dortmund.de}
\keywords{computational complexity, communication complexity.}
\begin{abstract}
Here we prove an asymptotically optimal lower bound on the information
complexity of the $k$-party disjointness function with the unique
intersection promise, an important special case of the well known
disjointness problem, and the $\AND_k$-function in the number in the hand model.
Our $\Omega(n/k)$ bound for disjointness improves on an
earlier $\Omega(n/(k \log k))$ bound by Chakrabarti {\it et al.}~(2003),
who obtained an asymptotically tight lower bound for one-way protocols,
but failed to do so for the general case.
Our result eliminates both the gap between the upper and the lower bound
for unrestricted protocols and the gap between the lower bounds for one-way
protocols and unrestricted protocols.\\

\end{abstract}
\maketitle

\section{Introduction}
Primarily, communication complexity, introduced by Yao~\cite{yao:1979},
deals with the amount of communication that is needed in distributed
computation, but apart from distributed computation, nowadays communication
complexity has found applications in virtually all fields of complexity
theory. The book by Kushilevitz and Nisan~\cite{kushilevitz/nisan:1997}
gives a comprehensive introduction to communication complexity
and its applications.

Suppose that $k$ players, each of them knowing exactly
one argument of a function $f(x_1,\dots,x_k)$ with $k$ arguments,
want to evaluate the function for the input that is distributed among them.
Clearly, to succeed at this task the players need to communicate.
Here we consider the case that the players communicate by writing to
a blackboard that is shared by all players. The rules that determine
who writes which message to the blackboard are usually called
a {\em protocol}. The protocol terminates if the value of the function
can be inferred from the contents of the blackboard, the
so-called {\em transcript} of the protocol.
Then the communication complexity of the function is
the minimum number of bits that the players need to write to the blackboard
in the worst case to jointly compute the result.
This setting is usually called the {\em number in the hand model}
since each part of the input is exclusively known to a single player
who figuratively hides the input in his hand.
In the randomized version of this model each player has access
to a private source of unbiased independent random bits and his
actions may depend on his input and his random bits.
For a {\em randomized $\eps$-error protocol} the output of the protocol
may be different from the value of the function $f$ with probability
at most $\eps$. The $\eps$-error randomized communication complexity
of a function is defined in the obvious way. A formal definition of $k$-party
protocols can be found in~\cite{kushilevitz/nisan:1997}.
Note that there are also other models of multi-party communication,
but these models are not the topic of this paper.

In recent publications~\cite{chakrabarti/others:2001,bar-yossef/others:2002,bar-yossef/others:2004,chakrabarti/others:2003}
lower bounds on the communication complexity of functions have been obtained
by using information theoretical methods. In this context communication
complexity is supplemented by an information theoretical counterpart,
the information complexity of a function. Roughly, the information
complexity of a function $f$ is the minimal amount of information that the
transcript of a protocol for $f$ must reveal about the input. Besides being a
lower bound for the communication complexity, information complexity
has additional nice properties with respect to so-called direct sum problems.

\subsection{Our Result}
In this paper we will prove an asymptotically optimal lower bound on
the communication complexity of the multi-party set disjointness problem
with the unique intersection promise.
\begin{definition}
In the $k$-party set disjointness problem each of the players is
given the characteristic vector of a subset of an $n$-element set.
It is promised that the subsets are either pairwise disjoint or that
there is a single element that is contained in all subsets and that
the subsets are disjoint otherwise. The players have to distinguish these
two cases, the output of a protocol for set disjointness should be $0$
in the first case and $1$ in the second case. If the promise is broken,
then the players may give an arbitrary answer.
\end{definition}
Here we will prove the following result about the randomized communication
complexity of the multi-party set disjointness problem in the
number in the hand model.
\begin{theorem}\label{thm:main-1}
For every sufficiently small constant $\eps>0$ the randomized $\eps$-error
communication complexity of the $k$-party set disjointness problem
with the unique intersection promise is bounded from below by $\Omega(n/k)$.
\end{theorem}
By the upper bound shown in~\cite{chakrabarti/others:2003} this result is
asymptotically optimal with respect to the number of players $k$ and
the size of the inputs $n$.
An important application of this problem is the proof of a lower bound
for the memory requirements of certain data stream
algorithms~\cite{alon/matias/szegedy:1999}. Our improvement
of the lower bound for disjointness does not have a significant impact
on this application. But we think that the disjointness problem is interesting
and important on its own since it is a well-known basic problem
in communication complexity theory~\cite{alon/matias/szegedy:1999,
bar-yossef/others:2004, chakrabarti/others:2003,kushilevitz/nisan:1997}.
Up to now the best known lower bound was $\Omega(n/(k\log{}k))$
by Chakrabarti, Khot, and Sun~\cite{chakrabarti/others:2003},
who also proved an asymptotically optimal lower bound for one-way protocols.
This result left a gap both between the upper and the lower bound
and between the lower bounds for one-way protocols and unrestricted protocols.
Our result closes these gaps.

Like the earlier results, our lower bound is based on an information
theoretical approach. The main ingredient of this approach is
a lower bound on the information complexity of the $\AND_k$-function,
the Boolean conjunction of $k$ bits.
Since Theorem~\ref{thm:main-1} will be a simple corollary
of this result, and more importantly, since $\AND_k$ is a basic building block
of any computation, the lower bound on the information complexity of $\AND_k$
is the main result of this paper. We postpone the precise statement
of this result to Theorem~\ref{thm:main-2} in Section~\ref{sec:and}
because some preparing definitions are needed beforehand.
But we stress here that our result also closes the gap between the upper and
lower bound on the conditional information complexity of $\AND_k$
for unrestricted protocols and the gap between the lower bounds
on the information complexity of $\AND_k$ for one-way protocols and
unrestricted protocols that was left open in~\cite{chakrabarti/others:2003}.

\subsection{Related Work}
The general disjointness problem without the unique intersection
promise has a long history in communication complexity theory.
Here we focus only on recent results for the multi-party set disjointness
problem with the unique intersection promise, and especially on lower
bounds that rely on information complexity arguments.
For older results we refer the reader to the book by Kushilevitz
and Nisan~\cite{kushilevitz/nisan:1997} and the references therein.

Alon, Matias, and Szegedy~\cite{alon/matias/szegedy:1999} proved
an $\Omega(n/k^4)$ lower bound for multi-party set disjointness and
applied this bound to prove lower bounds for the memory requirements
of data stream algorithms. Bar-Yossef, Jayram, Kumar, and
Sivakumar~\cite{bar-yossef/others:2004} improved this to a lower bound
of $\Omega(n/k^2)$. They introduced the direct sum approach on which later
results, including our result, are based and proved that the
information complexity of $\AND_k$ is bounded from below by $\Omega(1/k^2)$.
Chakrabarti, Khot, and Sun~\cite{chakrabarti/others:2003} improved
the lower bound for the information complexity of $\AND_k$
to $\Omega(1/(k\log{}k))$ and thereby improved the lower bound
for multi-party set disjointness to $\Omega(n/(k\log{}k))$.
They also proved an asymptotically optimal lower bound for
one-way protocols, a restricted model in which the players
communicate in a predetermined order.
Our result improves on these results, but furthermore we think that
our proof technique is a useful contribution to the framework for which
Bar-Yossef {\it et al.}~\cite{bar-yossef/others:2004} coined
the term ``information statistics''. Bar-Yossef {\it et al.} use
this term for the combination of information theory and other
statistical metrics on probability spaces.
We use the direct sum approach from~\cite{bar-yossef/others:2004},
but instead of the Hellinger distance that is used
in~\cite{bar-yossef/others:2004} we use the Kullback Leibler distance.
Since the Kullback Leibler distance is closely related to mutual information,
we do not loose precision in the transition from information theory
to statistical distance measures. By this, we are able to prove sharper bounds.
Like Chakrabarti {\it et al.}~\cite{chakrabarti/others:2003},
we take a closer look at the analytical properties of the functions
that are involved. Our improvements on this result are also due
to the fact that our Kullback Leibler distance based arguments are
very close to the information theory domain.

\section{Preliminaries}
\subsection{Notation}
We use lower case letters for constants and variables and upper
case letters for random variables. If the random variables $X$
and $Y$ have the same distribution, we briefly write ${X\sim{}Y}$.
For vector-valued variables we use a boldface font.
For example, $\vec{X}=(X_1,\dots,X_k)$ is a random vector whose
components are the random variables $X_i$ for $i=1,\dots,k$.
In this case let $\vec{X}_{-i}=(X_1,\dots,X_{i-1},X_{i+1},\dots,X_k)$
denote the vector $\vec{X}$ without the $i$th component.
A boldface zero $\vec{0}$ and boldface one $\vec{1}$ denote
the all-zero vector and all-one vector of appropriate size, respectively.
Thus $\vec{X}_{-i}=\vec{0}$ says that $X_j=0$ for
all $j\in\{1,\dots,k\}-\{i\}$.
For sums like $\sum_{i=0}^n a_i$ we sometimes do not explicitly
specify the bounds of summation and just write $\sum_{i} a_i$.
In this case the sum is taken over the set of all values of $i$
for which $a_i$ is meaningful. This set must be derived from context.
For example, the sum $\sum_{v} f(\Pr\{X\eq{}v\})$ should
be taken over all values $v$ in the range of $X$.
All logarithms, denoted by $\log$, are with respect to base $2$.

\subsection{Information Theory}\label{subsec:infotheory}
Here we can merely define our notation for the basic quantities from
information theory and cite some results that are needed in this paper.
For a proper introduction to information theory we refer the reader to the
book by Cover and Thomas~\cite{cover/thomas:1991}.
In the following let $h_2$ denote the binary entropy
function $h_2(p)=-p\log{}p-(1-p)\log{}(1-p)$ for $p\in[0,1]$.
Let $X$, $Y$, and $Z$ be random variables and let $E$ be an event, for example
the event ${Y=y}$. Then $\HH(X)$ denotes the entropy of
the random variable $X$ and $\HH(X|E)$ denotes the entropy of $X$ with
respect to the conditional distribution of $X$ given that
the event $E$ occurred. If there are several events separated by commas,
then we analogously use the conditional distribution of $X$ given that
all of the events occurred. Let $\HH(X|Y)$ denote the conditional entropy
of $X$ given $Y$. Recall that $\HH(X|Y)=\sum_{y}\Pr\{Y\eq{}y\}\HH(X|Y\eq{}y)$.
If we condition on several variables, we separate the variables by commas.
If we mix events and variables in the condition, we first list the variables,
after that we list the events, for example $\HH(X|Y,Z\eq{}z)$.
The mutual information of $X$ and $Y$ is $\II(X\col{}Y)=\HH(X)-\HH(X|Y)$
and $\II(X\col{}Y|E)=\HH(X|E)-\HH(X|Y,E)$ is the mutual information
of $X$ and $Y$ with respect to the conditional distribution of $X$ and $Y$
given that the event $E$ occurred.
The conditional mutual information of $X$ and $Y$ given $Z$
is $\II(X\col{}Y|Z)=\HH(X|Z)-\HH(X|Y,Z)$.
Recall that $\II(X\col{}Y|Z)=\sum_{z}\Pr\{Z\eq{}z\}\II(X\col{}Y|Z\eq{}z)$.

Suppose that the random variables $X$ and $Y$ have the same range.
Then the Kullback Leibler distance of their distributions
is $\DD(X,Y)=\sum_{v}\Pr\{X\eq{}v\}\log\frac{\Pr\{X\eq{}v\}}{\Pr\{Y\eq{}v\}}$.
If ${\Pr\{X\eq{}v\}=0}$ in the above sum, then the corresponding term
is $0$ independently of the value of $\Pr\{Y\eq{}v\}$, by continuity arguments.
If $\Pr\{X\eq{}v\}\neq{}0$ and $\Pr\{Y\eq{}v\}=0$ for some $v$,
then the whole sum is defined to be equal to $\infty$.
If $E$ is an event, then $(X|E)$ denotes the conditional distribution
of $X$ given that the event $E$ occurred, for example $\DD((X|E),X)$ is
the Kullback Leibler distance of the conditional distribution of $X$
given that the event $E$ occurred and the distribution of $X$.
Recall that the mutual information of $X$ and $Y$ is the
Kullback Leibler distance of the joint distribution $(X,Y)$
and the product distribution of the marginal distributions:
\begin{equation*}
    \II(X\col{}Y)
    =
    \sum_{x,y}
        \Pr\{X\eq{}x,Y\eq{}y\}\cdot
        \log
            \frac
                {\Pr\{X\eq{}x,Y\eq{}y\}}
                {\Pr\{X\eq{}x\}\cdot\Pr\{Y\eq{}y\}}
    \; \text{.}
\end{equation*}
The following lemma is a useful tool for the proof of lower bounds
on the Kullback Leibler distance of distributions.
A proof of the log sum inequality can be found in~\cite{cover/thomas:1991}.
\begin{lemma}[Log sum inequality]\label{lem:log-sum}
For nonnegative numbers $a_i$ and $b_i$, where $i=1,\dots,n$,
\begin{equation*}
    \sum_i a_i \log \frac{a_i}{b_i}
    \geq
    \left(\sum_i a_i\right) \log \frac{\sum_i a_i}{\sum_i b_i}
    \;\text{.}
\end{equation*}
\end{lemma}

Suppose that the random variables $X$ and $Y$ have the same finite range $R$.
Then the total variation distance of their distributions
is $\VV(X,Y)=\half\sum_{v}|\Pr\{X\eq{}v\}-\Pr\{Y\eq{}v\}|$.
It is a well-known fact (see e.g.~\cite{gibbs/su:2002})
that $\VV(X,Y)=\max_{S\subseteq{}R} |\Pr\{X\in{}S\}-\Pr\{Y\in{}S\}|$.
The following lemma by Kullback relates the Kullback Leibler distance
of distributions to their total variation distance.
\begin{lemma}[Kullback~\cite{kullback:1967}]\label{lem:kullback}
Suppose that $X$ and $Y$ are random variables that have the same finite range.
Then $\DD(X,Y)\geq{}2\cdot\VV(X,Y)^2$.
\end{lemma}

\subsection{Information Complexity}\label{subsec:infocomplexity}

The notion of the information cost of a protocol was introduced
by Chakrabarti, Shi, Wirth, and Yao~\cite{chakrabarti/others:2001}.
The information cost of a randomized protocol is the mutual information
of the input and the transcript of the protocol.
Then the information complexity of a function can be defined in the
canonical way. Here we will use the conditional information complexity
of a function, a refinement that was introduced by Bar-Yossef, Jayram,
Kumar, and Sivakumar~\cite{bar-yossef/others:2004}.
\begin{definition}\label{def:infocomplexity}
Let $B$ be a set, let $f\colon{}B^k\longrightarrow{}\binset$ be a function,
and let $\vec{X}\in{}B^k$ and $D$ be random variables.
Suppose that $P$ is a randomized $k$-party protocol for $f$ and
that $M(\vec{X})$ is the transcript of $P$ for the input $\vec{X}$.
Then the conditional information cost of $P$ with respect to $\vec{X}$
and $D$ is defined by
\begin{equation*}
    \icost(P;\vec{X}|D) = \II(M(\vec{X})\col{}\vec{X}|D)
    \;\text{.}
\end{equation*}
The conditional $\eps$-error information complexity $\IC_\eps(f;\vec{X}|D)$
of $f$ w.r.t. $\vec{X}$ and $D$ is the minimal conditional information cost
of a communication protocol for $f(\vec{X})$ where the minimum is taken
over all randomized $\eps$-error protocols for $f$.
\end{definition}

The information complexity of a function is a lower bound for
the communication complexity. A proof of the next theorem can
be found in~\cite{bar-yossef/others:2004}.
\begin{theorem}\label{thm:cc-geq-ic}
Let $B$ be a set, let ${f\colon{}B^k\longrightarrow{}\binset}$ be a function,
and let $\vec{X}\in{}B^k$ and $D$ be random variables.
Then the $\eps$-error communication complexity of $f$ is
bounded from below by $\IC_\eps(f;\vec{X}|D)$.
\end{theorem}

\subsection{The Direct Sum Paradigm}\label{subsec:direct-sum}

Information complexity has very nice properties with respect to
direct sum problems. In this section we summarize the approach of Bar-Yossef,
Jayram, Kumar and Sivakumar~\cite{bar-yossef/others:2004} using a slightly
different terminology. We call a problem $f$ a direct sum problem if it
can be decomposed into simpler problems of smaller size.
\begin{definition}\label{def:direct-sum}
Let $f\colon{}(B^n)^k\longrightarrow\binset$ be a function
and let $\vec{x}_i=(x_{i,1},\dots,x_{i,n})\in{}B^n$ for $i=1,\dots,k$.
If there are functions $g\colon{}\binset^n\longrightarrow\binset$
and $h\colon{}B^k\longrightarrow\binset$ such that
\begin{equation*}
    f(\vec{x}_1,\dots,\vec{x}_k)
    =
    g\left(\,h(x_{1,1},x_{2,1},\dots,x_{k,1})\,,\,\dots\,,\,h(x_{1,n},x_{2,n},\dots,x_{k,n})\,\right)
\end{equation*}
then the function $f$ is called a \mbox{$g$-$h$-}direct sum.
\end{definition}
Here the goal is to express a lower bound on the conditional information
complexity of~$f$ in terms of the conditional information complexity of
the simpler function~$h$ and the parameter~$n$.
In order for this approach to work, the joint distribution of the
inputs of~$h$ and the condition must have certain properties.
As a first requirement, the condition must partition the distribution
of the inputs into product distributions.
\begin{definition}
Let~$B$ be a set and let $\vec{X}=(X_1,\dots,X_k)\in{}B^k$ and~$D$
be random variables. The variable~$D$ partitions~$\vec{X}$, if for every~$d$
in the support of~$D$ the conditional distribution $(\vec{X}|D\eq{}d)$ is the
product distribution of the distributions $(X_i|D\eq{}d)$ for~$i=1,\dots,k$.
\end{definition}
The function $f$ can be decomposed into instances of the function $h$
if the distribution of the inputs of $f$ satisfies our second requirement.
\begin{definition}\label{def:collapsing-distribution}
Let $B$ be a set, let $g\colon{}\binset^n\longrightarrow\binset$
and $h\colon{}B^k\longrightarrow\binset$ be functions, and
let $\vec{X}\in{}B^k$ be a random variable.
If for every $i\in\{1,\dots,n\}$, for every $a\in{}B^k$, and for
every $\vec{x}=(\vec{x}_1,\dots,\vec{x}_n)\in(B^k)^n$ such
that $\vec{x}_j\in\support(\vec{X})$ for all $j$ 
\begin{equation*}
    g\left(h(x_1),\dots,h(x_{i-1}),h(a),h(x_{i+1}),\dots,h(x_n)\right)
    =
    h(a)
\end{equation*}
then the distribution of $\vec{X}$ is called collapsing for $g$ and $h$.
\end{definition}
If these two requirements are met, then the conditional information complexity
of~$f$ can be expressed in terms of the conditional information complexity
of~$h$ and the parameter~$n$.
\begin{theorem}[Bar-Yossef {\it et al.}~\cite{bar-yossef/others:2004}]\label{thm:direct-sum}
Suppose that $f\colon{}(B^n)^k\longrightarrow\binset$ is a
\mbox{$g$-$h$-}direct sum and that $\vec{X}\in{}B^k$ and $D$
are random variables such that the distribution of $\vec{X}$ is collapsing
for $g$ and $h$ and $D$ partitions $\vec{X}$.
Let $\vec{Y}=(\vec{Y}_1,\dots,\vec{Y}_k)\in{}(B^n)^k$
and $\vec{E}\in\support(D)^n$ be random variables
and let $Y_i^j$ and $E^j$ denote the projection of $\vec{Y}_i$ and $\vec{E}$
to the $j$th coordinate, respectively.
If the random variables $\vec{V}_j=((Y_1^j,\dots,Y_k^j),E^j)$ for $j=1,\dots,n$ are
independent and ${\vec{V}_j\sim(\vec{X},D)}$ for all $j$,
then $\IC_\eps(f;\vec{Y}|\vec{E})\geq{}n\cdot\IC_\eps(h;\vec{X}|D)$.
\end{theorem}
This direct sum approach can be applied to the $k$-party set
disjointness problem.
\begin{observation}
Let $\AND_\ell$ and $\OR_\ell$ denote the Boolean
conjunction and disjunction of $\ell$ bits, respectively.
Then the $k$-party set disjointness problem is a
\mbox{$\OR_n$-$\AND_k$-}direct sum.
\end{observation}
Consequently, for the proof of Theorem~\ref{thm:main-1} it is sufficient
to prove a lower bound on the conditional  information complexity
of $\AND_k$ for a distribution that satisfies the requirements
of Theorem~\ref{thm:direct-sum} and, in addition, honors the unique
intersection promise. A distribution with these properties is defined
in the following section.
This approach was already used in~\cite{bar-yossef/others:2004}
and~\cite{chakrabarti/others:2003}.

\section{The Information Complexity of \boldmath $\AND_k$}\label{sec:and}
For the following distribution of $D$ and the input $\vec{Z}=(Z_1,\dots,Z_k)$
of $\AND_k$ the variable $D$ partitions $\vec{Z}$ and the distribution
of $\vec{Z}$ is collapsing for $\OR_n$ and $\AND_k$.
Additionally, there is at most a single $i$ such that $Z_i=1$.
\begin{definition}\label{def:var-z}
From here on let $\vec{Z}=(Z_1,\dots,Z_k)\in\binset^k$ and $D\in\{1,\dots,k\}$
be random variables such that the joint distribution of $\vec{Z}$ and $D$
has the following properties: $D$ is uniformly distributed in $\{1,\dots,k\}$.
For all $i\in\{1,\dots,k\}$ we have $\Pr\{Z_j\eq{}0|D\eq{}i\}=1$ for $j\neq{}i$
and $\Pr\{Z_i\eq{}0|D\eq{}i\}=\Pr\{Z_i\eq{}1|D\eq{}i\}=\half$.
\end{definition}
Now we can state the main result of this paper, an asymptotically
optimal lower bound on the information complexity of the $AND_k$-function
for inputs that are distributed according to the last definition.
\begin{theorem}\label{thm:main-2}
Let $\eps<\frac{3}{10}\left(\,1-\sqrt{\half\log\frac{4}{3}}\,\right)$
be a constant. Then there is a constant $c(\eps)>0$ that does only depend
on $\eps$ such that $\IC_\eps(\AND_k;\vec{Z}|D)\geq{}c(\eps)/k$.
\end{theorem}
It is easy to see that $\icost(P;\vec{Z}|D)=1/k$ for a trivial deterministic
protocol $P$ for $\AND_k$ where each player in turn writes his input to
the blackboard until the first $0$ is written. Therefore our lower bound
is optimal. As we have seen, this result immediately implies
Theorem~\ref{thm:main-1}, the other main result of this paper.
In the rest of the paper we will outline the proof of Theorem~\ref{thm:main-2}.

\subsection{Some Basic Observations}
We start with some basic observations about the joint distribution
of the inputs and the transcript of a protocol for $\AND_k$ with
independent, uniformly distributed inputs. 
\begin{definition}
From now on, let $P$ be a fixed randomized $k$-player protocol
that computes $\AND_k$ with error at most $\eps$ and
for $\vec{x}\in\binset^k$ let $M(\vec{x})$ denote the transcript of $P$
for the input $\vec{x}$. Let $\vec{X}=(X_1,\dots,X_k)$ be a random variable
that is uniformly distributed in $\binset^k$ and let $T=M(\vec{X})$ denote
the transcript of $P$ for the the input $\vec{X}$.
\end{definition}
Note that the transcript $M(\vec{x})$ does depend on $\vec{x}$ and
the random inputs of the players. Thus even for a fixed input $\vec{x}$
the transcript is a random variable whose value depends on
the random bits used in the protocol.

A randomized $k$-party protocol can be seen as a deterministic
protocol in which the $i$th player has two inputs: The input to
the randomized protocol, in our case $X_i$, and as a second input
the random bits that are used by the $i$th player.
Then the first observation is a restatement of the fact that the
set of the inputs (real inputs and random bits) that correspond to a fixed
transcript is a combinatorial rectangle (see~\cite{kushilevitz/nisan:1997}
for a definition of combinatorial rectangles).
\begin{observation}[\cite{bar-yossef/others:2004,chakrabarti/others:2003}]\label{obs:cond-indep-1}
Let $\vec{x}=(x_1,\dots,x_k)\in\binset^k$ and let $t$ be an element
from the support of $T$. Then
$
    \Pr\{\vec{X}\eq{}\vec{x}|T\eq{}t\}
    =
    \prod_{i} \Pr\{X_i\eq{}x_i|T\eq{}t\}
$.
\end{observation}
We omit the simple combinatorial proof of this observation because
this basic property of $k$-party protocols was already used
in~\cite{bar-yossef/others:2004} and~\cite{chakrabarti/others:2003}.
The following observation is an immediate, but very useful consequence
of the previous one.
\begin{observation}\label{obs:cond-indep-2}
Let $\vec{x}=(x_1,\dots,x_k)\in\binset^k$ and let $t$ be an
element from the support of $T$.
Then $\Pr\{X_i\eq{}x_i|T\eq{}t,\vec{X}_{-i}\eq\vec{x}_{-i}\}=\Pr\{X_i\eq{}x_i|T\eq{}t\}$
for all $i\in\{1,\dots,k\}$.
\end{observation}
\proof
This observation follows immediately from Observation~\ref{obs:cond-indep-1}:
By adding the equality from Observation~\ref{obs:cond-indep-1}
for $(x_1,\dots,x_{i-1},0,x_{i+1},\dots,x_k)$
and $(x_1,\dots,x_{i-1},1,x_{i+1},\dots,x_k)$ we obtain
\begin{equation*}
    \Pr\{\vec{X}_{-i}\eq{}\vec{x}_{-i}|T\eq{}t\}
    =
    \prod_{j\neq{}i} \Pr\{X_j\eq{}x_j|T\eq{}t\}
    \;\text{.}
\end{equation*}
Using this and Observation~\ref{obs:cond-indep-1} verbatim yields
\begin{align*}
    \Pr\{X_i\eq{}x_i|T\eq{}t,\vec{X}_{-i}\eq\vec{x}_{-i}\}
    &=
    \frac
        {\Pr\{X_i\eq{}x_i,\vec{X}_{-i}\eq\vec{x}_{-i}|T\eq{}t\}}
        {\Pr\{\vec{X}_{-i}\eq\vec{x}_{-i}|T\eq{}t\}}\\
    &=
    \frac
        {\prod_j\Pr\{X_j\eq{}x_j|T\eq{}t\}}
        {\prod_{j\neq{}i}\Pr\{X_j\eq{}x_j|T\eq{}t\}}
    =
    \Pr\{X_i\eq{}x_i|T\eq{}t\}
    \; \text{.}
\end{align*}
\qed
The next observation relates the joint distribution
of $Z_i$ and $M(\vec{Z})$ given that $D=i$ to the joint distribution
of $X_i$ and $T=M(\vec{X})$ given that $\vec{X}_{-i}\eq\vec{0}$. Combined with
the previous observations, this will be the basis for the proof of
the main result.
\begin{observation}\label{obs:z-and-x}
Let $i\in\{1,\dots,k\}$.
Then $\II(M(\vec{Z})\col{}Z_i|D\eq{}i)=\II(T\col{}X_i|\vec{X}_{-i}\eq\vec{0})$.
\end{observation}
\proof
First observe that
$\Pr\{\vec{Z}\eq\vec{v},M(\vec{Z})=t|D\eq{}i\}=\Pr\{\vec{X}\eq\vec{v},T\eq{}t|\vec{X}_{-i}\eq\vec{0}\}$
for every $v\in\binset^k$ and every $t$ in the support of $M(X)$ and $M(Z)$.
This follows from the fact that the conditional distribution of $\vec{X}$
given that $\vec{X}_{-i}\eq\vec{0}$ is the same as the conditional
distribution of $\vec{Z}$ given that $D\eq{}i$, the fact that
the random inputs of $P$ are independent of $\vec{X}$ and $\vec{Z}$, and
the fact that the transcript is a function of the inputs and the random inputs.
Then the claim of the lemma is an immediate consequence of
the initial observation.
\qed

\subsection{Main Idea of the Proof}

Like the approach of Bar-Yossef {\it et al.}~\cite{bar-yossef/others:2004},
our approach is based on the observation that the distribution of the
transcripts of a randomized protocol for $\AND_k$ with small error must at
least be very different for the inputs $\vec{X}=\vec{0}$ and $\vec{X}=\vec{1}$.
The difference is expressed using some appropriate metric on probability
spaces. Then, by using Observations~\ref{obs:cond-indep-1}
and~\ref{obs:cond-indep-2}, this result is decomposed into results
about the distributions of $(X_i,M(\vec{X})|\vec{X}_{-i}\eq{}\vec{0})$
which are finally used to bound the conditional mutual information
of $\vec{Z}$  and $M(\vec{Z})$ given $D$ by using
Observation~\ref{obs:z-and-x}.
The result from~\cite{bar-yossef/others:2004} mainly uses the Hellinger
distance (see~\cite{gibbs/su:2002}) to carry out this very rough outline
of the proof. We will stick to the rough outline, but our result will
use the Kullback Leibler distance instead of the Hellinger distance.
Due to the limited space in the STACS-proceedings we can only present
proof-sketches of the technical lemmas in this section.
A version of this paper with full proofs can be found on the authors
homepage~\footnote{\tt http://ls2-www.cs.uni-dortmund.de/\textasciitilde{}gronemeier/}.

We will first decompose the Kullback Leibler distance of the
distributions $(T|\vec{X}\eq\vec{0})$ and $(T|\vec{X}\eq\vec{1})$ into
results about the joint distributions of $X_i$ and $T$ for $i=1,\dots,k$.
The result will be expressed in terms of the following function.
\begin{definition}\label{def:g}
From now on, let $g(x)=x\log \frac{x}{1-x}$.
\end{definition}
Note that the left hand side of the equation in the following lemma
is the Kullback Leibler distance of $(T|\vec{X}\eq\vec{0})$
and $(T|\vec{X}\eq\vec{1})$ if $S$ is the set of all possible transcripts.
\begin{lemma}\label{lem:idea-2}
Let $S$ be a subset of the set of all possible transcripts. Then
\begin{equation*}
    \sum_{t\in{S}} \Pr\{T\eq{}t|\vec{X}\eq\vec{0}\}\cdot
        \log\frac{\Pr\{T\eq{}t|\vec{X}\eq\vec{0}\}}{\Pr\{T\eq{}t|\vec{X}\eq\vec{1}\}}
    =
    2\sum_i \sum_{t\in{}S} \Pr\{T\eq{}t|\vec{X}_{-i}\eq\vec{0}\}\cdot
        g(\Pr\{X_i\eq{}0|T\eq{}t\})
    \;\text{.}
\end{equation*}
\end{lemma}
\proofsketch
The proof of this lemma is mainly based on the fact that
\begin{equation*}
    \frac{\Pr\{T\eq{}t|\vec{X}\eq\vec{0}\}}{\Pr\{T\eq{}t|\vec{X}\eq\vec{1}\}}
    =
    \frac{\Pr\{\vec{X}\eq\vec{0}|T\eq{}t\}}{\Pr\{\vec{X}\eq\vec{1}|T\eq{}t\}}
    \;\text{.}
\end{equation*}
Then Observation~\ref{obs:cond-indep-1} can be applied to decompose
the $\log$-function into a sum. Finally, we use that
$\Pr\{T\eq{}t|\vec{X}\eq\vec{0}\}=
2\Pr\{T\eq{}t|\vec{X}_{-i}\eq\vec{0}\}\cdot\Pr\{X_i\eq{}0|T\eq{}t\}$
by Observation~\ref{obs:cond-indep-2}.
\qed

Next, we will express a lower bound on $\II(M(\vec{Z})\col{}Z|D)$ in terms
of the following function $f$ and set $B(\alpha)$.
\begin{definition}\label{def:f}
From now on, let $f(x)=x\log 2x + \frac{1-x}{2}\log 2(1-x)$.
\end{definition}
\begin{definition}\label{def:set-b-alpha}
Let $B(\alpha)$ denotes the set of all transcripts $t$ such that
${\Pr\{X_i\eq{}0|T\eq{}t\}<\alpha}$ for all $i\in{}\{1,\dots,k\}$.
\end{definition}
The role of the parameter $\alpha$ will become apparent later.
The only property that is needed for the proof of the following lemma
is that $\alpha>1/2$.
\begin{lemma}\label{lem:idea-1}
Let $\alpha>\half$ be a constant. Then
\begin{equation*}
    \II(M(\vec{Z})\col{}\vec{Z}|D)
    \geq
    \frac{1}{k} \sum_i \sum_{t\in{}B(\alpha)}
        \Pr\{T\eq{}t|\vec{X}_{-i}\eq\vec{0}\} \cdot f(\Pr\{X_i\eq{}0|T\eq{}t\})
    \;\text{.}
\end{equation*}
\end{lemma}
\proofsketch
This lemma can be proved by using that $f(x)=\half(f_1(x)+f_2(x))$
where $f_1(x)= x\log 2x + (1-x)\log 2(1-x)$ and $f_2(x)= x \log 2x$.
It is sufficient to prove that the lower bound holds for $f_1$ and $f_2$
instead of $f$. To this end one can show that
\begin{equation*}
    \II(M(\vec{Z})\col{}\vec{Z}|D)
    =
    \frac{1}{k}\sum_i\sum_{t}
        \Pr\{T\eq{}t|\vec{X}_{-i}\eq\vec{0}\}
        \cdot{}f_1(\Pr\{X_i\eq{}0|T\eq{}t\})
    \;\text{.}
\end{equation*}
Then the bound for $f_1$ is obvious since $f_1(x)$ is nonnegative
for all $x\in[0,1]$. The bound for $f_2$ use the fact
that $f_1(x)=f_2(x)+f_2(1-x)$, that $f_2(x)\geq{}0$ for $x\in[1/2,1]$,
and that
\begin{equation*}
    \sum_{t}\Pr\{T\eq{}t|\vec{X}_{-i}\eq\vec{0}\}
    \cdot{}f_2(\Pr\{X_i\eq{}1|T\eq{}t\})
\end{equation*}
is nonnegative.
\qed

The right hand sides of the equation in Lemma~\ref{lem:idea-2}
and the inequality in Lemma~\ref{lem:idea-1} look very similar.
In fact, if there was a positive constant $c$
such that $c\cdot{}f(x)\geq{}g(x)$ for all $x\in[0,1]$,
then for a complete proof of Theorem~\ref{thm:main-2}
it would be sufficient to show that the Kullback Leibler distance
of $(T|\vec{X}\eq\vec{0})$ and $(T|\vec{X}\eq\vec{1})$ is bounded from below
by a constant $c(\eps)$ if the error of the protocol $P$ is bounded by $\eps$.
Unfortunately $f(x)\leq{}1$ for $x\in[0,1]$ while $g(x)$ is not
bounded from above for $x\in[0,1]$. So this naive first idea does not work.
But the function $g(x)$ is bounded in every interval $[0,\beta]$ where $\beta<1$.
The following Lemma shows that we can easily bound $f(x)$ from below in terms
of $g(x)$ if we restrict $x$ to an appropriate interval $[0,\beta]$.
\begin{lemma}\label{lem:comparing-f-g}
There is a constant $\beta>\frac{1}{2}$ such that
$4\cdot{}f(x)\geq{}g(x)$ for all $x\in[0,\beta]$.
\end{lemma}
This lemma can probably be proved in many ways.
By inspection and numeric computations it is easy to verify that
it holds for $\beta\approx0.829$.
Here it is more important to note that our choice of the function $f$
is one of the crucial points of our proof: The function $g(x)$ is
negative for $x\in[0,\half)$ and nonnegative and increasing 
for $x\in[\half,1]$. Furthermore $g(\half)=0$ and in the interval $[\half,1]$
the slope of $g(x)$ is bounded from below by a positive constant.
It will become clear in Lemma~\ref{lem:unbiased-2} that we have
to lower bound $f(x)$ in terms of $g(x)$ for $x\approx\half+O(\frac{1}{k})$
where $k$ is the number of players. Recall that $f(x)=\half(f_1(x)+f_2(x))$
where $f_1(x)= x\log 2x + (1-x)\log 2(1-x)$ and $f_2(x)= x \log 2x$
and that we prove Lemma~\ref{lem:idea-1} by lower bounding the
mutual information of $M(\vec{Z})$ and $\vec{Z}$ in terms of $f_1(x)$ and $f_2(x)$.
Thus $f_1(x)$ and $f_2(x)$ would be natural candidates for the
function $f(x)$. Unfortunately, neither $f_1(x)$ nor $f_2(x)$ alone
does work in our proof. The function $f_1(x)$ is nonnegative
for $x\in[0,1]$, therefore $f_1(x)\geq{}g(x)$ for $x\in[0,\half]$,
but the slope of $f_1(x)$ is too small in the interval $[\half,1]$.
It turns out that $f_1(\half+\frac{1}{k})\approx1/k^2$.
If we used the function $f_1(x)$ instead of $f(x)$ in our proof,
we could only obtain an $\Omega(1/k^2)$ lower bound for the information
complexity of $\AND_k$. The function $f_2(x)$ does not suffer from this
problem since the slope of $f_2(x)$ in $[\half,1]$ is bounded from below
by a constant. But here we have the problem that $f_2(x)$ is too small
for $x\in[0,\half)$. For every constant $c>0$ such
that $c\cdot{}f(x)\geq{}g(x)$ in the interval $x\in[\half,1]$
we have $g(x)>c\cdot{}f(x)$ in the interval $x\in[0,\half)$.
Luckily, for the average $f(x)$ of $f_1(x)$ and $f_2(x)$ the good properties
of the functions are preserved while the bad properties ``cancel out''.
The bounded slope for $x\in[\half,1]$  of $f(x)$ is inherited from $f_2(x)$.
The fact that $f(x)$ is not to small for $x\in[0,\half)$ is inherited
from $f_1(x)$.

We can use the set $B(\alpha)$ in Lemma~\ref{lem:idea-1} and the set $S$ 
in Lemma~\ref{lem:idea-2} to restrict $t$ to the transcripts
that satisfy $\Pr\{X_i\eq{}0|T\eq{}t\}\leq{}\beta$ for all $i\in\{1,\dots,k\}$.
Then, by our previous observations, it is easy to lower bound
$f(\Pr\{X_i\eq{}0|T\eq{}t\})$ in terms of $g(\Pr\{X_i\eq{}0|T\eq{}t\})$.
\begin{definition}\label{def:set-b}
Let $\beta$ be the constant from Lemma~\ref{lem:comparing-f-g}.
recall that $B(\alpha)$ denotes the the set of all transcripts $t$ such that
$\Pr\{X_i\eq{}0|T\eq{}t\}<\alpha$ for all $i\in{}\{1,\dots,k\}$.
Then $B$ is a shorthand notation for the set $B(\beta)$.
\end{definition}
Unfortunately, the restriction of $t$ to the set $S=B$ complicates the proof
of a lower bound for the left hand sum in Lemma~\ref{lem:idea-2} since
we remove the largest terms from the sum.
For example, we will see in the proof of Corollary~\ref{cor:zero-error} that
for zero-error protocols the set $B$ does only contain transcripts for
the output $1$. Therefore, by the zero-error property,
$\Pr\{T\in{}B|\vec{X}\eq\vec{0}\}=0$ for zero error protocols and
the left hand sum in Lemma~\ref{lem:idea-2} is equal to $0$.
Consequently, without further assumptions that do not hold in general it is
impossible to prove large lower bounds on the sum in Lemma~\ref{lem:idea-2}
for the set $S=B$. However, the next Lemma shows that we can lower bound
the sum, if we assume that $\Pr\{T\in{}B|\vec{X}\eq\vec{0}\}$
is sufficiently large.
\begin{lemma}\label{lem:unbiased-2}
Suppose that $\Pr\{T\in{}B|\vec{X}\eq\vec{0}\}\geq\frac{3}{4}$ and
that the error $\eps$ of the protocol $P$ is bounded
by $\eps<\frac{3}{10}\left(\,1-\sqrt{\half\log\frac{4}{3}}\,\right)$. Then
\begin{equation*}
    \sum_{t\in{}B} \frac{\Pr\{T\eq{}t|\vec{X}\eq\vec{0}\}}{\Pr\{T\in{}B|\vec{X}\eq\vec{0}\}}
        \cdot \log\frac{\Pr\{T\eq{}t|\vec{X}\eq\vec{0}\}}{\Pr\{T\eq{}t|\vec{X}\eq\vec{1}\}}\\
    \geq
    \min\left\{
        \log\frac{3}{2},
        2\left(1-\frac{10}{3}\eps\right)^2-\log\frac{4}{3}
    \right\} > 0
    \;\text{.}
\end{equation*}
\end{lemma}
\proofsketch
For the proof of this lemma we consider two cases:
If $\Pr\{T\in{}B|\vec{X}\eq\vec{1}\}<\frac{1}{2}$ then we can
use the log sum inequality (Lemma~\ref{lem:log-sum}) to lower bound
the sum on the left hand side.
If $\Pr\{T\in{}B|\vec{X}\eq\vec{1}\}\geq\frac{1}{2}$ then the error
of the protocol $P$ under the condition that $T\in{}B$ must be small
both for the input $\vec{X}=\vec{0}$ and the input $\vec{X}=\vec{1}$.
With this assumption we can lower bound the left hand side using
Lemma~\ref{lem:kullback} since in this case the total variation
distance of $(T|\vec{X}\eq\vec{0},T\in{}B)$
and $(T|\vec{X}\eq\vec{1},T\in{}B)$ is large.
\qed
Note that, by Lemma~\ref{lem:idea-2} and the fact that the slope of $g(x)$
is bounded from below by a positive constant for $x\in[1/2,1]$, this lower
bound can be met if $\Pr\{X_i\eq{}0|T\eq{t}\}=\half+\Theta(\frac{1}{k})$
for all $i\in\{1,\dots,k\}$ and every $t\in{}B$.

By Lemma~\ref{lem:unbiased-2}, under the condition
that $\Pr\{T\in{}B|\vec{X}\eq\vec{0}\}\geq\frac{3}{4}$
our initial naive plan of bounding $f$ in terms of $g$ does work.
The details of this idea are elaborated on in the proof of
Theorem~\ref{thm:main-3}. Next, we look at the case
that $\Pr\{T\in{}B|\vec{X}\eq\vec{0}\}$ is small.
It turns out that this assumption alone already leads to
a large lower bound on $\II(M(\vec{Z})\col{}\vec{Z}|D)$.
\begin{lemma}\label{lem:biased}
Let $\alpha$ be a constant subject to $1/2<\alpha\leq{}1$. Then
\begin{equation*}
    \II(M(\vec{Z})\col{}\vec{Z}|D)
    \geq
    \frac{1}{2k}\cdot\Pr\{T\notin{}B(\alpha)|\vec{X}\eq\vec{0}\}\cdot(1-h_2(\alpha))
    \text{.}
\end{equation*}
\end{lemma}
\proofsketch
The proof of this lemma is based on the fact that, by the definition
of $B(\alpha)$, under the condition that $T=t\notin{}B(\alpha)$
the entropy of $X_i$ is bounded by $h_2(\alpha)<1$ for at least one $i$.
\qed

Now all prerequisites for a full proof of Theorem~\ref{thm:main-2}
are in place. It is implied by the following theorem because $P$ was assumed
to be an arbitrary $\eps$-error protocol for $\AND_k$.
\begin{theorem}\label{thm:main-3}
Let $\eps<\frac{3}{10}\left(\,1-\sqrt{\half\log\frac{4}{3}}\,\right)$
be a constant. If the error of the protocol $P$ is bounded by $\eps$, then
there is a constant $c(\eps)>0$ that does only depend on $\eps$ such that
\begin{equation*}
    \II(M(\vec{Z})\col{}\vec{Z}|D)\geq{}\frac{c(\eps)}{k}
    \text{.}
\end{equation*}
\end{theorem}
\proof
Recall that $B$ is the set of all transcripts $t$ such that
${\Pr\{X_i\eq{}0|T\eq{}t\}<\beta}$ for all $i\in{}\{1,\dots,k\}$,
where $\beta$ is the constant from Lemma~\ref{lem:comparing-f-g}.
For the proof of the lemma we will consider two cases.

For the first case, assume that $\Pr\{T\in{}B|\vec{X}\eq\vec{0}\}\leq\frac{3}{4}$.
In this case we can apply Lemma~\ref{lem:biased} with $\alpha=\beta$
and we get
\begin{equation*}
    \II(M(\vec{Z})\col{}\vec{Z}|D)
    \geq
    \frac{1}{2k}\Pr\{T\notin{}B|\vec{X}\eq\vec{0}\}(1-h_2(\beta))\\
    \geq
    \frac{1}{8k}(1-h_2(\beta))
    \;\text{.}
\end{equation*}
Note that in this case the lower bound does not depend on $\eps$
and that, since $\beta>1/2$, there is a constant $c_1>0$ such that the right
hand side of the last inequality is bounded from below by $c_1/k$.

For the second case, assume that $\Pr\{T\in{}B|\vec{X}\eq\vec{0}\}>\frac{3}{4}$.
In this case we first apply Lemma~\ref{lem:idea-1} for $\alpha=\beta$,
thus $B(\alpha)=B$, then Lemma~\ref{lem:comparing-f-g}, and finally
Lemma~\ref{lem:idea-2} for the subset $S=B$ to get
\begin{align*}
    \II(M(\vec{Z})\col{}\vec{Z}|D)
    &\geq
    \frac{1}{k} \sum_i \sum_{t\in{}B}
        \Pr\{T\eq{}t|\vec{X}_{-i}\eq\vec{0}\} \cdot f(\Pr\{X_i\eq{}0|T\eq{}t\})\\
    &\geq
    \frac{1}{4k} \sum_i \sum_{t\in{}B}
        \Pr\{T\eq{}t|\vec{X}_{-i}\eq\vec{0}\} \cdot g(\Pr\{X_i\eq{}0|T\eq{}t\})\\
    &=
    \frac{1}{8k}
    \sum_{t\in{B}} \Pr\{T\eq{}t|\vec{X}\eq\vec{0}\}\cdot
        \log\frac{\Pr\{T\eq{}t|\vec{X}\eq\vec{0}\}}{\Pr\{T\eq{}t|\vec{X}\eq\vec{1}\}}
    \;\text{.}
\end{align*}
Then, by the assumption $\Pr\{T\in{}B|\vec{X}\eq\vec{0}\}>\frac{3}{4}$, we can
apply Lemma~\ref{lem:unbiased-2} to obtain
\begin{align*}
    \II(M(\vec{Z})\col{}\vec{Z}|D)
    &\geq
    \frac{1}{8k}\cdot
    \Pr\{T\in{}B|\vec{X}\eq\vec{0}\}\cdot
    \min\left\{
        \log\frac{3}{2},
        2\left(1-\frac{10}{3}\eps\right)^2-\log\frac{4}{3}
    \right\}\\
    &\geq
    \frac{3}{32k}\cdot
    \min\left\{
        \log\frac{3}{2},
        2\left(1-\frac{10}{3}\eps\right)^2-\log\frac{4}{3}
    \right\}
    \;\text{.}
\end{align*}
For $\eps<\frac{3}{10}\left(\,1-\sqrt{\half\log\frac{4}{3}}\,\right)$
the minimum in the last inequality is a positive constant that does only
depend on the constant $\eps$. Hence, there is a constant $c_2(\eps)>0$
that does only depend on the constant $\eps$ such that the right hand side
is bounded from below by $c_2(\eps)/k$. The claim of the Lemma follows
from the two cases if we choose $c(\eps)=\min\{c_1,c_2(\eps)\}$.
\qed

\subsection{A Simple Lower Bound for Zero-Error Protocols}

For zero-error protocols a lower bound can be proved by using
only Lemma~\ref{lem:biased}.
\begin{corollary}\label{cor:zero-error}
For every randomized $k$-player zero-error protocol with input $\vec{Z}$ and
transcript $M(\vec{Z})$ the conditional information cost satisfies
$\II(M(\vec{Z})\col{}\vec{Z}|D)\geq 1/(2 k)$.
\end{corollary}
\proof
Consider the transcript $T$ of the protocol $P$ for the input $\vec{X}$.
Then the corollary follows immediately from Lemma~\ref{lem:biased}
if we set $\alpha=1$:
Recall that the output of the protocol can be inferred from the transcript
and let $P(t)$ denote the output of the protocol $P$ for transcript $t$.
Suppose that $P(t)=0$. Then $\Pr\{X_i\eq{}0|T\eq{}t\}=1$ for at least one $i$ since otherwise,
by Observation~\ref{obs:cond-indep-1}, $\Pr\{\vec{X}\eq\vec{1}|T\eq{}t\}>0$
and under the condition $T\eq{}t$ the output of $P$ would be wrong with a
nonzero probability. Clearly this is not possible for zero-error protocols,
hence $\Pr\{T\notin{}B(1)|P(T)\eq{}0\}=1$.
Under the condition $\vec{X}\eq{}\vec{0}$ the output of $P$ is $0$ with
probability $1$, again by the zero-error property, therefore the last
observation implies that $\Pr\{T\notin{}B(1)|\vec{X}\eq{}\vec{0}\}=1$ and
obviously $1-h_2(1)=1$.
\qed

\section*{Acknowledgments}
Thanks to Martin Sauerhoff for helpful discussions and proofreading.



\begin{thebibliography}{10}

\bibitem{alon/matias/szegedy:1999}
N.~Alon, Y.~Matias, and M.~Szegedy.
\newblock The space complexity of approximating the frequency moments.
\newblock {\em J. Comput. Syst. Sci.}, 58(1):137--147, 1999.

\bibitem{bar-yossef/others:2002}
Z.~Bar-Yossef, T.~S. Jayram, R.~Kumar, and D.~Sivakumar.
\newblock Information theory methods in communication complexity.
\newblock In {\em Proc. of 17th CCC}, pages 93--102, 2002.

\bibitem{bar-yossef/others:2004}
Z.~Bar-Yossef, T.~S. Jayram, R.~Kumar, and D.~Sivakumar.
\newblock An information statistics approach to data stream and communication
  complexity.
\newblock {\em J. Comput. Syst. Sci.}, 68(4):702--732, 2004.

\bibitem{chakrabarti/others:2003}
A.~Chakrabarti, S.~Khot, and X.~Sun.
\newblock Near-optimal lower bounds on the multi-party communication complexity
  of set disjointness.
\newblock In {\em Proc. of 18th CCC}, pages 107--117, 2003.

\bibitem{chakrabarti/others:2001}
A.~Chakrabarti, Y.~Shi, A.~Wirth, and A.~C. Yao.
\newblock Informational complexity and the direct sum problem for simultaneous
  message complexity.
\newblock In {\em Proc. of 42nd FOCS}, pages 270--278, 2001.

\bibitem{cover/thomas:1991}
T.~M. Cover and J.~A. Thomas.
\newblock {\em Elements of Information Theory}.
\newblock Wiley-Interscience, 1991.

\bibitem{gibbs/su:2002}
A.~L. Gibbs and F.~E. Su.
\newblock On choosing and bounding probability metrics.
\newblock {\em International Statistical Review}, 70:419, 2002.

\bibitem{kullback:1967}
S.~Kullback.
\newblock A lower bound for discrimination information in terms of variation.
\newblock {\em IEEE Trans. Inform. Theory}, 4:126--127, 1967.

\bibitem{kushilevitz/nisan:1997}
E.~Kushilevitz and N.~Nisan.
\newblock {\em Communication Complexity}.
\newblock Cambridge University Press, 1997.

\bibitem{yao:1979}
A.~C. Yao.
\newblock Some complexity questions related to distributive computing
  (preliminary report).
\newblock In {\em Proc. of 11th STOC}, pages 209--213, 1979.

\end{thebibliography}

\end{document}